\documentclass[12pt,twocolumn,apj]{emulateapj}
\usepackage[utf8]{inputenc}
\usepackage{amsmath}
\usepackage{amsfonts}
\usepackage{amssymb}
\usepackage{graphicx}
\usepackage{graphics}
\usepackage{mathrsfs}
\usepackage{color}

\begin{document}
\title{The effect of vertical temperature gradient on the propagation of three-dimensional waves in a protoplanetary disk}
\author{Wing-Kit Lee\altaffilmark{*} and Pin-Gao Gu}
\affil{Academia Sinica, Institute of Astronomy and Astrophysics}
\altaffiltext{*}{wklee@asiaa.sinica.edu.tw}

\begin{abstract}
Excitation and propagation of waves in a thermally stratified disk with an arbitrary vertical temperature profile are studied. Previous analytical studies of three-dimensional waves had been focused on either isothermal or polytropic vertical disk structures. However, at the location in a protoplanetary disk where the dominant heating source is stellar irradiation, the temperature gradient may become positive in the vertical direction. We extend the analysis to study the effects of the vertical temperature structure on the waves that are excited at the Lindblad resonances. For a hotter disk atmosphere, the $f$-mode contributes less to the torque and remains confined near the midplane as it propagates away from the resonances. On the other hand, the excitation of the $g$-modes is stronger. As they propagate, they channel to the top of disk atmosphere and their group velocities decrease. The differences compared to previous studies may have implications in understanding the wave dynamics in a realistic disk structure.
\end{abstract}

\section{Introduction}

When an embedded planet orbits around its central star, the tidal interactions between the protoplanetary disk and the planet will lead to excitation of waves from resonance locations \citep{1979ApJ...233..857G}. These waves propagate radially and may eventually damp or dissipate through shocks and transfer angular momentum to the disk. Such process may also lead to radial migration of the planet \citep{1996Natur.380..606L, 1997Icar..126..261W,2002ApJ...572..566R} and change of disk structure such as gap formation \citep[e.g.,][]{1996ApJ...460..832T, 2006Icar..181..587C}. Therefore, the properties of these hydrodynamic waves and their relation to the disk structure are crucial to the understanding of the planet/disk dynamics. 

Since the disks are cold (i.e., the sound speed is much less than the circular speed), they are geometrically thin (i.e., the scale height is less than radial distance to the star, $H < r$). Yet, the vertical thermal structure is still important to explain the observed spectral energy distribution of the disks \citep[e.g.,][]{1997ApJ...490..368C, 1998ApJ...500..411D}. Measurement of such vertical structure in a protoplanetary disk has been made possible previously with single-dish telescopes \citep[e.g.,][]{akiyama2011thermal} and more recently with ALMA \citep{2013ApJ...774...16R}. On the other hand, recent near-infrared polarization intensity maps suggest that the surfaces of some protoplanetary disks are not smooth on the large scale but exhibit concave spiral features \citep{2014ApJ...795...71T}. However, the origins of these surface features are still not well-understood as the surface density enhancement required would be too large in the 2D linear theories \citep[e.g.,][]{2015MNRAS.451.1147J}. As a result, three-dimensional wave theories may shed light on the relation between the planet-disk interaction and observations \citep[e.g.,][]{2015ApJ...809L...5D, 2015arXiv150703599Z}.

Three-dimensional wave propagation in accretion disks has been studied by several authors in the past. For analytical convenience, these studies assumed a disk that is stable against convection, with either an isothermal vertical structure \citep{1993ApJ...409..360L} or a polytropic one \citep{1995MNRAS.272..618K, 1998ApJ...504..983L}. Since a polytropic disk contains a finite boundary where gas density and pressure vanish above it, such model is considered as a wave-guide model due to the boundary conditions applied at the midplane (usually a parity condition) and the surface (e.g., vanishing Lagrangian pressure perturbation). A set of discrete modes can be identified by the vertical eigenfunctions and the corresponding dispersion relation. \citet{1998MNRAS.297..291O} provided a generalized analysis with magnetic field and a concise classification of waves in a disk, namely, $f$-mode (fundamental), $p$-modes (acoustic), $r$-modes (inertial), and $g$-modes (internal gravity). As the waves are excited at resonance locations where the wavelength is large \citep{1979ApJ...233..857G}, only $f$-, $g$-, and $r$-modes are launched at the Lindblad resonances (LRs) where the Doppler-shifted forcing frequency equals the epicyclic frequency. \citet{1998ApJ...504..983L} studied the wave excitation of these modes at the LRs and found that most of the angular momentum are indeed carried by the $f$-mode. The authors also found that the wave is evanescent in the vertical direction and experiences wave-channeling, which is contrary to the previous results that wave refraction occurs due to the gradient of sound speed \citep{1990ApJ...364..326L}. \citet{1999ApJ...515..767O} considered a polytropic disk with an isothermal atmosphere. They found that the wave angular momentum concentrates near the transition between the two layers while propagating radially. Subsequently, \citet{2002MNRAS.332..575B} studied the nonlinear propagation of axisymmetric waves using hydrodynamic simulation and verified the wave-channeling of these waves.

While these studies provide a general theory of wave excitation and propagation, the cases for a hotter disk atmosphere are not explored. In particular, in previous models, the temperature decreases along the normal direction to the disk as the disk becomes less optically thick and the heat can be easily carried away by radiation. However, at the outer part of the disk, the stellar irradiation may be very efficient to generate a hotter layer as the disk surface is flared.

To study the effects of the vertical temperature gradient, we adopt the aforementioned wave-guide model. The background disk structure is assumed to be slowly varying in the radial direction (except at LRs) with respect to the waves. Appropriate boundary conditions at the midplane and at the top of the disk atmosphere are applied to construct a boundary eigenvalue problem, with the Doppler-shifted frequency of the wave being the eigenvalue. The vertical temperature profile is parameterized and modeled such that the temperature varies with height from the midplane and gradually reaches a steady (isothermal) atmosphere which is heated by the stellar irradiation from the central star. The properties of the waves, such as wave-channeling, can be studied by varying the parameters of the temperature profile. In our model, a vertical hydrostatic equilibrium is assumed to be consistent with the prescribed temperature profile. As we focus on the wave propagation, the change of thermal properties of the disk (e.g., variation in the adiabatic index $\gamma$) at the top of the atmosphere and the consequence of wave dissipation are neglected in this work.

The paper is organized as follows. In Section \ref{sec:formulation}, we first review the formulation and derive the governing equations and the relevant boundary conditions. In Section \ref{sec:results}, we present and discuss the result for the wave excitation and propagation. Finally, we summarize this work and draw a conclusion in Section \ref{sec:conclusion}.

\section{Formulation of the Problem}
\label{sec:formulation}
\subsection{Governing Equations}
A geometrically thin and cold gas disk orbiting around a central young stellar object is considered. An embedded planet is treated as a perturber and tidally interacts with the disk. For simplicity, the self-gravity of the gas is neglected although it may be important in the very early stage of the disk. The basic state of the system is described by an axisymmetric disk in a time-steady equilibrium without accretion. The inviscid hydrodynamic equations reads
\begin{align}
\label{eq:full0_1}
\frac{\partial \rho}{\partial t} + \nabla\cdot( \rho \mathbf{u}) &= 0, \\
\label{eq:full0_2}
\frac{\partial \mathbf{u}}{\partial t} + \mathbf{u}\cdot\nabla \mathbf{u} &= -\frac{1}{\rho}\nabla P - \nabla \Phi_{\rm tot},
\end{align}
where $\rho$, $\mathbf{u}$, and $P$ are the gas density, velocity, and pressure, respectively, and $\Phi_{\rm tot}$ is the total gravitational potential. In this work, we use the cylindrical coordinates $(r,\varphi,z)$ centered at the star. The total potential $\Phi_{\rm tot}$ is a sum of the contributions from the central star (subscript s) and the orbiting planet (subscript p):
\begin{align}
\Phi_{\rm tot} = \Phi_{\rm s}(r,z) + \Phi_{\rm p}(r,\varphi,z,t),
\end{align}
where $\Phi_{\rm p}$ includes an indirect term due to the choice of the origin being away from the center of mass. To close the equations, we model the basic state of the disk with an empirical formula for the temperature and assume the linear perturbation is adiabatic. The effects of the magnetic field and viscosity are also ignored in this work.

\subsection{Basic State}
The basic state of the disk can be solved by considering the radial and vertical directions separately. The equilibrium velocity is given by $\mathbf{u}_0 = r\Omega(r) \mathbf{e}_\varphi$, where the angular frequency of the gas $\Omega(r)$ is determined by the radial force balance at the midplane,
\begin{align}
\label{eq:eq_radial}
r\Omega(r)^2 = \frac{\partial \Phi_{\rm s}}{\partial r}\Big|_{z=0} + \frac{1}{\rho_0}\frac{\partial P_0}{\partial r}\Big|_{z=0},
\end{align}
where $\rho_0$ and $P_0$ are the equilibrium density and pressure, respectively. The gravitational potential of the central star is given by
\begin{align}
\Phi_{\rm s}(r,z) = -\frac{GM}{\sqrt{r^2+z^2}},
\end{align}
where $M$ is the mass of the central star. For a cold disk, the small correction due to the pressure force and the gravity of the planet are ignored, so that the equilibrium disk is Keplerian (i.e., $\Omega \propto r^{-3/2}$). The vertical disk structure is determined by the vertical hydrostatic equilibrium,
\begin{align}
\label{eq:eq_vert}
\frac{\partial P_0}{\partial z} = - \rho_0 g_z,
\end{align}
where $g_z = \Omega(r)^2 z$ is the $z$-component of the acceleration due to stellar gravity in the thin-disk limit.

Assuming the ideal gas law applies, we have
\begin{align}
P_0(r,z) = \frac{k_B}{m_a}\rho_0(r,z) T_0(r,z),
\end{align}
where $k_B$ is the Boltzmann's constant, $m_a$ is the average molecular mass (in grams), $\rho_0(r,z)$ and $T_0(r,z)$ are the equilibrium gas density and temperature, respectively. The local scale height $H$ at the midplane is defined in terms of the midplane temperature; that is
\begin{align}\label{eq:midplanescaleheight}
H^2 = \frac{c^2}{\Omega^2} = \frac{k_B T_{\rm mid}}{m_a \Omega^2},
\end{align}
where $c$ and $T_{\rm mid} = T(r,z=0)$ are the midplane sound speed and temperature, respectively. For a cold disk, the vertical extent of interest is of order of the scale height, but much smaller than the radius,
\begin{align}
\frac{z}{r} \sim \frac{H(r)}{r} = \epsilon \ll 1.
\end{align}
For $H/r = 0.05$ at LRs, the error of $g_z$ due to the thin-disk approximation is about 0.24 at the top boundary (see Section \ref{sec:boundary}).

\subsection{Model for Vertical Structure}
\label{sec:modelvertical}
Consistent with the assumption of a cold and thin disk, we further assume the separation of variables applies such that the radial variation of the equilibrium variables (e.g., $\rho_0$, $P_0$) are determined by their midplane values (e.g., $\rho_{\rm mid}$, $P_{\rm mid}$) which are labeled with subscript ``mid''. The midplane variables are varying in a radial length scale that is larger than $H$. Therefore, it is useful to define a dimensionless $z$ coordinate $Z=z/H(r)$, and
\begin{align}
T &= \frac{T_0(r,Z)}{T_{\rm mid}(r)}, \\
P &= \frac{P_0(r,Z)}{P_{\rm mid}(r)}, \\
\rho &= \frac{\rho_0(r,Z)}{\rho_{\rm mid}(r)},
\end{align}
where $T$, $P$, and $\rho$ are dimensionless counterparts of $T_0$, $P_0$, and $\rho_0$, respectively, and are functions of $Z$. The vertical hydrostatic equilibrium in Equation \eqref{eq:eq_vert} can be written in the following dimensionless form,
\begin{align}\label{eq:hydroeq}
\frac{dP}{d Z} = - \frac{PZ}{T},
\end{align}
where $P = \rho T$. Given a vertical temperature profile $T(Z)$, the equilibrium pressure and density can be obtained by integrating Equation \eqref{eq:hydroeq} and using the ideal gas law.

\subsubsection{Temperature Profiles}
\label{sec:tempprofile}

\begin{figure}
\includegraphics[width=0.5\textwidth]{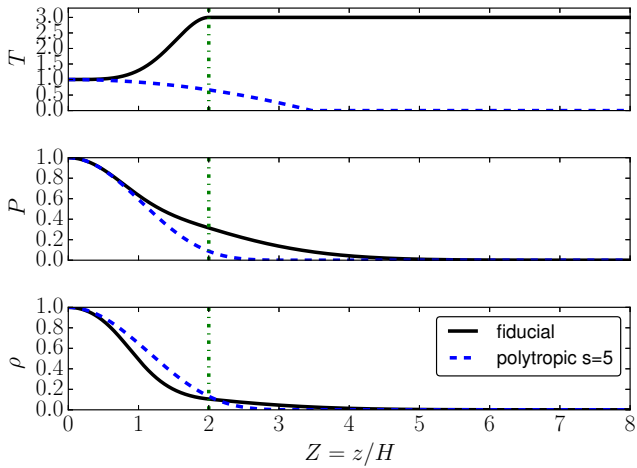}
\caption{Vertical profiles of the equilibrium state. The top, middle, and bottom panels are the dimensionless profiles for gas temperature, pressure, and density, respectively. The solid line shows the fiducial model with $A=2.0$, $Z_{\rm a} = 2.0$ (indicated by the vertical dashed-dotted line). The dashed line shows the polytropic disk with $s=5$. We denote $H$ as the scale height at the midplane.}
\label{fig:backgroundstate}
\end{figure}

In the outer region of the disk, the effect on the temperature due to stellar irradiation dominates viscous heating. In addition, long-wavelength emissions from embedded large-sized grains and some molecules become optically thin to the disk. Therefore, it is expected that the temperature of the disk increases with height from the midplane. For simplicity, we do not explicitly solve the energy equation with sources of heating and cooling. This allows us to focus on the linear response to the disk. Instead, we parameterize the vertical structure using the relative temperature increase in the disk atmosphere, i.e., $A = (T_{\rm atm} - T_{\rm mid})/T_{\rm mid}$. Inspired by \citet{2002A&A...389..464D} and \citet{2013ApJ...774...16R}, our temperature model gradually extends into an atmosphere with constant temperature, which is given by
\begin{align}
\label{eq:temperature1}
T(Z) = 
\begin{cases} 
1 + (1/2)A\left[1-\cos\pi \left(Z/Z_{\rm a}\right)^n\right] & \mbox{if } Z < Z_{\rm a}\\
1 + A &\mbox{if } Z \geq Z_{\rm a}, \\ 
\end{cases} 
\end{align}
where $A$ is the fractional change of the vertical temperature, $Z_{\rm a}$ is the transition height to the isothermal atmosphere at a temperature $T=1+A$, and $n$ is a parameter determining how rapid is the transition. The parameter $n=2$ is chosen such that the equilium can be solved analytically (Appendix \ref{sec:appendixB}) and there is a simple stability criterion against convection as we will discuss below. The transition from the bottom layer to the isothermal atmosphere is smooth such that $T'(Z_{\rm a})=0$, where the prime (and hereafter) denotes the $Z$ derivative. Above $Z_{\rm a}$, both gas pressure and density fall off as a Gaussian with distance from the midplane (i.e., $\propto \exp{-Z^2/(1+A)}$). 

\subsection{Deviation from Equilibrium}
We proceed to derive a set of equations for the variables deviated from the basic state. The velocity perturbation can be expressed as $\mathbf{u}_1 = u \mathbf{e}_r + {\rm v} \mathbf{e}_\varphi + {\rm w} \mathbf{e}_z$, where $u=u_{r1}$, ${\rm v}=u_{\varphi 1}$, and ${\rm w}=u_{z 1}$. Keeping the nonlinear terms and canceling the terms for the equilibrium disk, the component form of Equations \eqref{eq:full0_1}-\eqref{eq:full0_2} can be rewritten as
\begin{align}
\label{eq:full2_1}
&\frac{\partial \rho}{\partial t} + \frac{1}{r}\frac{\partial}{\partial r} \left(r \rho u \right) + \frac{1}{r}\frac{\partial}{\partial \varphi}(\rho {\rm v}) + \frac{\partial}{\partial z} (\rho {\rm w}) = 0,
\end{align}
\begin{align}
\label{eq:full2_2}
\frac{D u}{Dt} - \frac{{\rm v}^2}{r} - 2\Omega {\rm v} &= -\frac{1}{\rho}\frac{\partial P_1}{\partial r} -\frac{\partial}{\partial r}\Phi_{\rm p},\\
\label{eq:full2_3}
\frac{D {\rm v}}{Dt} +\frac{u {\rm v}}{r} + 2 B u &= -\frac{1}{\rho r}\frac{\partial P_1}{\partial \varphi}-\frac{1}{r}\frac{\partial}{\partial \varphi}\Phi_{\rm p},\\
\label{eq:full2_4}
\frac{D {\rm w}}{Dt} &= -\frac{1}{\rho}\frac{\partial P_1}{\partial z} -\frac{\partial}{\partial z}\Phi_{\rm p},
\end{align}
where
\begin{align}
\frac{D}{Dt} = \frac{\partial}{\partial t} + u \frac{\partial}{\partial r} + \frac{u_\varphi}{r}\frac{\partial}{\partial \varphi} + {\rm w} \frac{\partial}{\partial z}
\end{align}
is the Lagrangian derivative, $u_\varphi = r\Omega(r) + {\rm v}$ is the total circular velocity, $P_1$ is the pressure perturbation, $\Phi_{\rm p}$ is the potential perturbation, and the Oort's constant $B$ is defined by
\begin{align}
B = \frac{r}{2}\frac{d\Omega}{dr} + \Omega.
\end{align}
The coefficient $2B$ in Equation \eqref{eq:full2_3} is also commonly expressed as $\kappa^2/2\Omega$, where $\kappa$ is the epicyclic frequency defined by $\kappa^2 = 4B\Omega$. In deriving Equations \eqref{eq:full2_1}-\eqref{eq:full2_4}, no assumption has been made for the amplitude of the perturbation.

\subsection{Linear Perturbation}
\label{sec:linpert}
We proceed to linearize Equations \eqref{eq:full2_1}-\eqref{eq:full2_4} by considering the linear waves. Assuming the embedded planet is orbiting circularly at an angular frequency $\Omega_{\rm p}$, the planet's potential can be expressed as a Fourier series that is periodic in $\varphi$ and $t$,
\begin{align}
\Phi_{\rm p} = \sum_m \Phi_m(r,z) \exp[ im(\varphi-\Omega_{\rm p} t)],
\end{align}
where $m$ is a positive integer. As Equations \eqref{eq:full2_1}-\eqref{eq:full2_4} do not depend on $t$ and $\varphi$ explicitly, they are linearized and Fourier-transformed into the following form
\begin{align}
\label{eq:full3_1}
&-i\hat{\omega}\rho_1 + \frac{1}{r}\frac{\partial}{\partial r}(r\rho_0 u) + \frac{im\rho_0{\rm v}}{ r} + \frac{\partial }{\partial z}(\rho_0 {\rm w}) = 0, \\
\label{eq:full3_2}
&-i\hat{\omega}u - 2\Omega {\rm v} + \frac{1}{\rho_0}\frac{\partial}{\partial r}P_1 = - \frac{\partial}{\partial r}\Phi_m, \\
\label{eq:full3_3}
&-i\hat{\omega}{\rm v} + 2B u + \frac{im}{\rho_0 r}P_1 = -\frac{im}{r} \Phi_m,\\
\label{eq:full3_4}
&-i\hat{\omega}{\rm w} + g\frac{\rho_1}{\rho_0} + \frac{1}{\rho}\frac{\partial}{\partial z}P_1 = - \frac{\partial}{\partial z}\Phi_m,
\end{align}
where $\hat{\omega}(r)=m[\Omega_{\rm p}-\Omega(r)]$, and the subscripts of the Fourier components for $\rho_1$, $u$, ${\rm v}$, ${\rm w}$ and $P_1$ are suppressed for clarity. The perturbational equation for the specific entropy reads
\begin{align}
\label{eq:full3_5}
&-i\hat{\omega}\left(\frac{P_1}{P_0}-\frac{\gamma \rho_1}{\rho_0}\right) + {\rm w} \frac{d}{d z} \ln \left(\frac{P_0}{\rho_0^\gamma}\right) = 0,
\end{align}
where $\gamma$ is the adiabatic index.

We focus on the horizontal resonances and ignore the vertical forcing due to the gravity of the planet. In the thin-disk limit, we ignore the terms due to the azimuthal gradient of ${\rm v}$ and $P_1$ in Equations \eqref{eq:full3_1} and \eqref{eq:full3_3}, respectively. By eliminating azimuthal velocity ${\rm v}$, Equations (\ref{eq:full3_2}) and (\ref{eq:full3_3}) can be combined into the following equation,
\begin{align}
\label{eq:full3_6}
(\hat{\omega}^2 - \kappa^2)u + \frac{i\hat{\omega}}{\rho_0}\frac{\partial P_1}{\partial r} = -i\hat{\omega}\frac{d\Phi_m}{dr} +\frac{2im\Omega}{r}\Phi_m,
\end{align}
where the forcing on the right-hand-side of the equation is only dominant at the LRs.

\subsection{Local Approximation and Normal Modes}
In order to study the vertical wave structure, we consider a radially restricted region such that the radial variation of the equilibrium state (subscript 0) can be neglected. This local approximation, or equivalently shearing-sheet approximation \citep{1965MNRAS.130..125G, 1993ApJ...406..596G}, allows the separation of variables in the radial and vertical directions in the governing equations. In particular, we assume the equilibrium state is a function of $z$ only and obtain a set of linear ordinary differential equations (ODEs) in $z$. We consider the waves excited at the LRs, where $\hat{\omega}(r=r_L) = \pm \kappa$.

\subsubsection{Near the Lindblad Resonances}
In the proximity of the LRs, the wavelength of the perturbation is comparable to radial extent of the forcing. Therefore, the usual plane wave approximation breaks down and we follow \citet{1998ApJ...504..983L} to expand the singular term to the linear order of dimensionless displacement $x=(r-r_L)/r_L$, that is
\begin{align}\label{eq:LR1}
\kappa^2 - \hat{\omega}^2 \simeq \mathscr{D} x,
\end{align}
where $\mathscr{D} = rd/dr(\kappa^2-\hat{\omega}^2)|_{r_L}$. For a Keplerian disk, $\mathscr{D}=-3(1\pm m)\Omega^2_L$, where $\Omega_L = \Omega(r=r_L)$ and $m>0$. Thus, Equation \eqref{eq:full3_6} becomes
\begin{align}\label{eq:radialforce}
-\mathscr{D} x u + \frac{i\hat{\omega}}{\rho_0}\frac{\partial P_1}{\partial r} = -i\hat{\omega}\frac{d\Phi_m}{dr} +\frac{2im\Omega}{r}\Phi_m.
\end{align}

\subsection{Free Waves}

We first consider the homogeneous solutions (i.e., free waves) of linearized equations. These eigenfunctions can be then used as the basis functions to expand the perturbational potential $\Phi_m$ \citep{1998ApJ...504..983L}. Each eigenfunction can be categorized into different modes by studying the corresponding dispersion relation $\hat{\omega}(k)$ where $k$ is the radial wavenumber. As the torque is proportional to $\Phi_m$, the fraction of each mode can be computed. To take into the account the linear term of $x$ in Equation \eqref{eq:LR1} in the proximity of the LRs, we make use of the following separation of variables suggested by \citet{1998ApJ...504..983L},
\begin{align}
\label{eq:sepvar1}
u(x,z) &= {\rm Ai} (qx) \tilde{u}(z),\\
\label{eq:sepvar2}
{\rm v}(x,z) &= i{\rm Ai} (qx) \tilde{\rm v}(z),\\
\label{eq:sepvar3}
{\rm w}(x,z) &= q{\rm Ai}' (qx) \tilde{\rm w}(z),\\
\label{eq:sepvar4}
P_1(x,z) &= iq{\rm Ai}' (qx) \tilde{P}_1(z),\\
\label{eq:sepvar5}
\rho_1(x,z) &= iq{\rm Ai}' (qx) \tilde{\rho}_1(z),
\end{align}
where $x$ is the dimensionless distance from the LR, $q$ is the effective radial wavenumber, and ${\rm Ai}(x)$ is the Airy function satisfying the following ODE,
\begin{align}
{\rm Ai}''(x) - x {\rm Ai}(x) = 0.
\end{align}
In the absence of forcing, Equations \eqref{eq:full3_1}-\eqref{eq:full3_5} can be combined and simplified into two ODEs by using Equations \eqref{eq:sepvar1}-\eqref{eq:sepvar5}. We define the following dimensionless variables
\begin{align}
\label{eq:dimless1}
X(Z) &= \hat{\omega}\tilde{P}_1(z)/(\Omega^3 H^2 \rho_0), \\
\label{eq:dimless2}
W(Z) &= \tilde{\rm w}(z)/(\Omega H), \\
\label{eq:dimless3}
F &= \hat{\omega}/\Omega, \\
\label{eq:dimless4}
Q &= (H/r)^{2/3} q, \\
\label{eq:dimless5}
S &= \mathscr{D}/\Omega^2.
\end{align}
After some algebra and making use of dimensionless variables, we obtain the following differential equations,
\begin{align}
\label{eq:dxdz}
\frac{dX}{dZ} &=\frac{N^2}{Z} X + \left[F^2 - N^2\right]W, \\
\label{eq:dwdz}
\frac{dW}{dZ} &= \left[\lambda - \frac{\rho}{\gamma P}\right] X - \frac{P'}{\gamma P} W,
\end{align}
where $F^2=1$ at LRs, $\lambda=Q^3/S$ is a unknown parameter, and
\begin{align}
\label{eq:dimless0}
N^2 = \frac{Z}{\gamma}\frac{d}{dZ}\ln{\frac{P}{\rho^\gamma}},
\end{align}
is the square of the dimensionless Brunt–V\"{a}is\"{a}l\"{a} frequency (normalized by $\Omega$).  The equilibrium variables are obtained by Equation \eqref{eq:hydroeq} in Section \ref{sec:modelvertical}. When the problem is solved with appropriate boundary conditions, the parameter $\lambda$ is determined as an eigenvalue. Equations \eqref{eq:dxdz} and \eqref{eq:dwdz} follow those obtained by \citet{1998ApJ...504..983L} except that we assume an arbitrary hydrostatic background profile instead of a polytropic disk. We rewrite the equations in a form that does not require solving Equation \eqref{eq:hydroeq} in advance, i.e.,
\begin{align}
\label{eq:dxdz_2}
\nonumber
\frac{dX}{dZ} &=\left[F^2 - \left(\frac{\gamma-1}{\gamma}\right)\frac{Z^2}{T} - \frac{ZT'}{T} \right] W \\
&+ \left[\left(\frac{\gamma-1}{\gamma}\right)\frac{Z}{T}+\frac{T'}{T}\right] X,\\
\label{eq:dwdz_2}
\frac{dW}{dZ} &= \left[\lambda - \frac{1}{\gamma T}\right] X + \frac{Z}{\gamma T} W,
\end{align}
where we have used
\begin{align}
N^2 = \left(\frac{\gamma-1}{\gamma}\right)\frac{Z^2}{T}+\frac{ZT'}{T}.
\end{align}
For non-decreasing temperature profiles ($A \geq 0$) in Equation \eqref{eq:temperature1} ($T'(Z)\geq 0$), the disk is stable against convection $N^2 \geq 0$. For $A < 0$ and $n=2$ in Equation \eqref{eq:temperature1}, the stability criterion against convection is $|A| \leq (\gamma-1)Z_{\rm a}^2/\gamma\pi \sim 5$ for our choice of parameters $n=2$ and $Z_{\rm a}=2$. Thus, the background is stable against convection in our explored parameter space. 

\subsection{Boundary Conditions}
\label{sec:boundary}
Equations \eqref{eq:dxdz_2}-\eqref{eq:dwdz_2} are the governing differential equations near the LRs. Away from the LRs ($F \neq 1$), the equations can be obtained by replacing $\lambda$ with $K^2 / (F^2-\kappa^2)$ (see Appendix \ref{sec:appendixA}; Equations \eqref{eq:dxdzNLR} and \eqref{eq:dwdzNLR}). They are numerically integrated and are subjected to boundary conditions at the midplane ($Z=0$) and at an arbitrary defined surface (at $Z=Z_{\rm b}$). As the system is symmetric in $Z$ in our simple model, the waves take one of the parities. We consider only the even solution where 
\begin{align}
\frac{dX}{dZ}\Big|_{Z=0} = 0 \quad\text{and}\quad W(Z=0)=0,
\end{align}
in which the pressure perturbation is an even function in $Z$ and the warping modes are ignored. 

The adiabatic index is assumed to be constant everywhere ($\gamma=5/3$), such that both the sound speed and buoyancy frequency are continuous across different layers. The upper boundary condition is applied such that the Lagrangian perturbation of pressure goes to zero at the top of the atmosphere $Z_{\rm b}$. In terms of the dimensionless variables, it reads
\begin{align}
X_{\rm b} - W_{\rm b} Z_{\rm b} = 0,
\end{align} 
where $X_{\rm b}$ and $W_{\rm b}$ are evaluated at $Z=Z_{\rm b}$. 

\section{Results and Discussion}
\label{sec:results}

A local model is constructed by assuming slow radial variation of the equilibrium state. For simplicity, the vertical equilibrium temperature profile is taken as a smooth continuous function.  The functional form is chosen to mimic the vertical structure determined by the radiative transfer model \citep[e.g.,][]{2002A&A...389..464D}. The temperature profile is characterized by three parameters, namely, $A$ the fractional increase of the temperature of the atmosphere relative to that of the midplane, $Z_{\rm a}$ the transition height to the atmosphere, and $Z_{\rm b}$ the upper boundary. In principle, the isothermal atmosphere extend to the infinity, in which the internal gravity ($g$) modes have a continuous spectrum. However, in order to study the behavior of each mode, we apply a upper boundary condition well above the transition height, such that the $f$-mode and $g$-modes appear to be discrete. For $r$-modes, its spectrum remains discrete as in the locally isothermal disk \citep{1993ApJ...409..360L}. In this work, we follow the convention to label each mode of the same type by the number of nodes in the pressure perturbation (i.e., $X$).

\begin{figure}[!htb]
\includegraphics[trim=0cm 2cm 0cm 2cm, clip=true, width=0.5\textwidth]{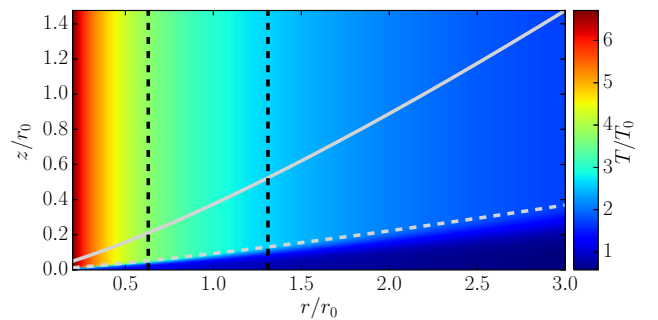}
\caption{Temperature profile for $A=2.0$. The horizontal and vertical axes are $r$ and $z$ in the unit of corotation radius $r_0$. The color scale shows the relative temperature with respect to midplane temperature $T_0$ at $r=r_0$. The dashed and solid gray lines are the transition location ($Z_a=2.0$) and upper computational boundary ($Z_b=8.0$), respectively. The vertical black dashed lines denote the LRs for $m=2$. Here we adopt a power-law $T_{\rm mid}(R)\propto R^{-1/2}$ and the aspect ratio $H/r=0.05$ at the outer LR.}
\label{fig:temperature2d}
\end{figure}

The background equilibrium density and pressure are obtained by solving the equations for hydrostatic equilibrium. The basic vertical structure is generally smoother than the polytropic disk counterpart (see Figure \ref{fig:backgroundstate}). A reference model for $A=2.0$ and $Z_{\rm a} = 2.0$ is used, as shown in Figure \ref{fig:temperature2d}. The parameters are varied in order to understand their effects on wave excitation and propagation.

At $A=0$, our model reduces to the locally isothermal disk with adiabatic response. This was previously studied by \citet{1993ApJ...409..360L}, in which a boundary condition is applied such that the wave energy density is finite at $z\rightarrow \infty$. In our case, since an artificial upper boundary applied, the discrete $g$-modes are recovered. However, $g$-modes are only weakly excited in this case.

\subsection{Torque Fraction}
\label{sec:torquefraction}

At the LRs, the torque due to the external potential can be evaluated as \citep{1979ApJ...233..857G, 1998ApJ...504..983L},
\begin{align}
\label{eq:r1}
\mathscr{T} = -{\rm sgn}(\hat{\omega})\frac{\pi^2 m \Sigma_L \Psi^2_m}{|\mathscr{D}|},
\end{align}
where $\Sigma_L = \int \rho dz$ is the column density at the LR, and 
\begin{align}
\label{eq:r2}
\Psi_m = r\frac{d\Phi_m}{dr}-\frac{2m\Omega}{\hat{\omega}}\Phi_m.
\end{align}
The formula from two-dimensional analysis is valid here because the vertical forcing of the planet is ignored. By projecting $\Psi$ onto the basis eigenfunctions, the fraction of the torque due to each mode of waves can be obtained. We denote $f_j$ the fraction of torque carried by each mode $j$ \citep{1998ApJ...504..983L}, which is given by
\begin{align}
\label{eq:r3}
f_j = \Bigg| \int \rho_0 X_j dZ \Bigg|^2 / \Sigma_L \int \rho_0 X_j^2 dZ,
\end{align}
where we used the relation between $\tilde{u}$ and dimensionless pressure perturbation $X$:
\begin{align}
\label{eq:r4}
\frac{\tilde{u}}{\Omega H} = -\lambda \left(\frac{r}{H}\right) X,
\end{align}
which is obtained from Equations \eqref{eq:radialforce} and \eqref{eq:dimless1}.

\subsubsection{Dependence on Temperature Gradient}

\begin{figure*}[!htb]
\centering
\includegraphics[width=\textwidth]{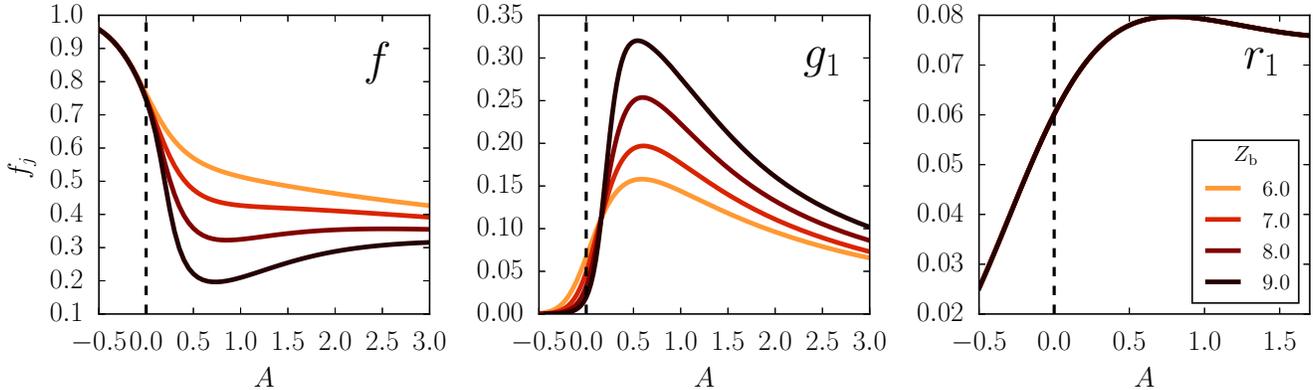}
\caption{Torque fractions of the $f$-mode (\emph{left}), $g_1$-mode (\emph{middle}), and $r_1$-mode (\emph{right}) versus temperature parameter $A$ for different values of the height of upper boundary $Z_{\rm b}$. The vertical dashed line indicates the locally isothermal disk ($A=0$). The transition height is at $Z_{\rm a} = 2.0$. Note that for $r_1$ mode, the torque fractions are the same for different $Z_{\rm b}$.}
\label{fig:torque-all}
\end{figure*}

For a simpler analysis, the dimensionless transition height $Z_{\rm a}$ is kept fixed at $2.0$. Therefore, the temperature gradient in the layer near the midplane is proportional to the temperature parameter $A$. The parameter changes both the background density $\rho_0$ and eigenfunctions $(X,W)$. As indicated in Equation \eqref{eq:r3}, the torque fraction is determined by the overlap integral between $\rho_0$ and $X$. Since $\sum_j f_j = 1$ among all modes, the torque fraction of any particular mode also depends on that of others. Most of the torque is carried by the first few lower-order modes (i.e., $f$, $r_1$ and $g_1$, where the subscript 1 means one node in $X$), although other higher-order modes are also excited when $A$ is large. Therefore, we focus on these lower-order modes for the analysis. The dependence of $f_j$ on $A$ for different values of $Z_{\rm b}$ is shown in Figure \ref{fig:torque-all}. 

For $f$-mode, the torque fraction increases with decreasing $A$ and becomes less sensitive to $Z_{\rm b}$. For $g_1$ mode, the torque fraction peaks at $A\sim 0.5$ and is generally larger for larger $Z_{\rm b}$. On the other hand, the $r_1$ mode contributes more torque than the $g_1$-mode when $A<0$. While its torque fraction also peaks at some finite value of $A$, the torque fraction is independent of $Z_{\rm b}$ (Figure \ref{fig:torque-all}).

The general increasing trend of $f_{f}$ for decreasing $A$ ($\lesssim 1$) can be explained by the decrease of $f_j$ for $g_1$- and $r_1$-modes. In particular, $\rho_0$ becomes more concentrated near the midplane and so the torque integral in Equation \eqref{eq:r3} depends sensitively on the overlapping of $\rho_0$ and $X$ for small $Z$. Either a smaller numerical value (e.g., $g_1$-mode) or an oscillatory solution (e.g., $r_1$-mode) can result in a small torque fraction. 

\subsubsection{Dependence on Upper Boundary}

Since $\rho_0$ is independent of the location of the upper boundary $Z_b$ in our model, the torque fraction depends only on the functional form of $X$. For $f$- and $g_1$-modes, the eigenfunctions $X$ are similar to those in a polytropic disk, where $X(Z)$ is generally an increasing function. While $X(Z_{\rm b})$ is a local maximum, the value of $X$ for $f$-mode at the midplane varies (without the effect of amplitude) and becomes very large at some frequency $F$. This eventually relates to the wave-channeling that we will discuss in the next section. Note that $X$ has zero and one node for $f$- and $g$-modes, respectively. Larger $Z_{\rm b}$ generally results in a smaller overlapping of $X$ and $\rho_0$ near the midplane. Thus, $f_j$ {\it decreases} with increasing $Z_{\rm b}$ for $f$-mode and $g_1$-mode (for small $A$). However, when $A$ is large, the amplitude of oscillation near the node for $g_1$-mode is also large. This cancels out the contribution in the torque integral and results in a reverse trend of $f_j$ with $Z_{\rm b}$. For $r_1$-mode, the torque fraction is independent of $Z_{\rm b}$. This is because $X$ is flat and close to zero for large $Z$ and so is not sensitive to $Z_{\rm b}$.

\subsection{Wave Propagation}

\begin{figure}[!ht]
\includegraphics[width=0.5\textwidth]{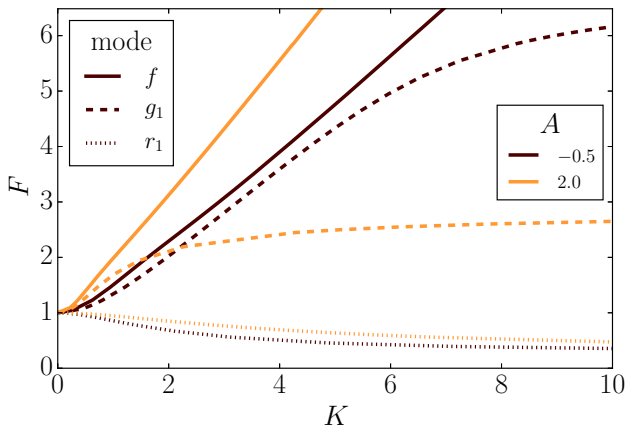}
\caption{Dispersion relation of different modes ($Z_{\rm a}=2$,$Z_{\rm b}=8$) are shown for $A=-0.5$ and 2.0. The solid, dashed, and dotted lines represent the $f$-, $g_1$- and $r_1$-modes, respectively. The waves are launched at LRs where $F=1$ and $K=0$.}
\label{fig:KF}
\end{figure}

The wave propagation of each mode is determined by the corresponding dimensionless dispersion relation $F(K)$, where $F=\hat{\omega}/\Omega$ and $K=kH$. In Figure \ref{fig:KF}, the dispersion relations are shown for $f$-, $g_1$-, and $r_1$-modes ($Z_{\rm a}=2$ and $Z_{\rm b}=8$). The slope of $F(K)$ for $f$-mode increases with larger $A$. It is expected for its acoustic character for $K \rightarrow 0$ \citep{1998ApJ...504..983L}, although there is no simple WKB relation for the moderate values of $K$. On the other hand, as $g$-modes are driven by buoyancy \citep{1998MNRAS.297..291O}, the forcing frequency approaches the Brunt–V\"{a}is\"{a}l\"{a} frequency near the top of the atmosphere for moderate values of $K$ such that $F(K) \sim N \propto 1/(1+A)^{1/2}$ (see Equation \eqref{eq:dimless0}). Using the temperature in the atmosphere, the ratio between $F$ for $A=-0.5$ and $2.0$ is about $\sqrt{6}\sim 2.4$, which is similar to the estimate from Figure \ref{fig:KF}. 

\begin{figure}[!ht]
\includegraphics[width=0.5\textwidth]{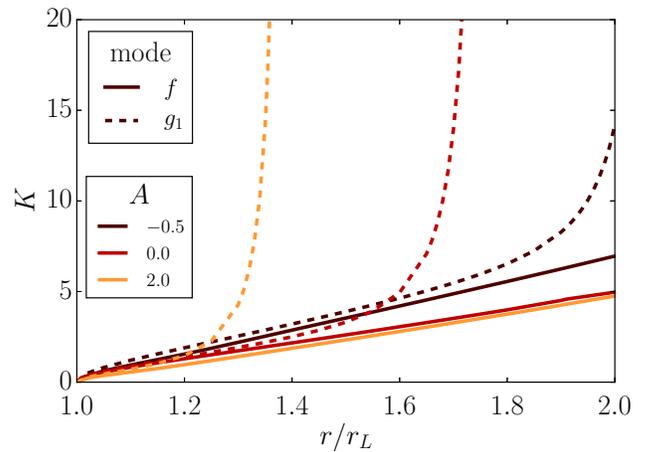}
\caption{Plot of radial wavenumber versus radius for $A=-0.5$, 0.0, and 2.0 ($Z_{\rm a}=2$,$Z_{\rm b}=8$). The solid and dashed lines represent the $f$- and $g_1$-modes, respectively.}
\label{fig:RK}
\end{figure}

As the Doppler-shifted frequency $\hat{\omega}$ is a function of $r$ for a given azimuthal wavenumber $m$, the radius can be expressed as
\begin{align}
\frac{r}{r_0}= (1 + F/m)^{2/3},
\end{align}
where $r_0$ is corotation radius. Note that $F$ is positive for $r>r_0$ and negative for $r<r_0$. The LRs are located at $r^\pm_L / r_c = (1 \pm 1/m)^{2/3}$, where the plus and minus signs indicate the outer and inner LRs, respectively. For the purpose for illustrating wave propagation, we consider only the outer LR ($r_L = r_L^+$). In Figure \ref{fig:RK}, we show the frequencies for $f$- and $g_1$-modes that propagate from the outer LR with $m=2$ (note that $r$-modes are absent in this region). In particular, the wavenumber $g_1$ mode rapidly increases as it propagates outward. As the group velocity of $g_1$-mode decreases significantly, the wave amplifies at some finite distance from the LRs ($r \leq 2 r_L$) due to the wave action conservation explained below. In particular, the location where $g_1$-modes ``piled-up" are closer to the LRs for the increasing $A$. For realistic disk with positive $A$ and higher $m$, we expect such effect occurs closer to the LRs. 

For waves that propagates in a non-uniform moving medium, the conservation of energy or angular momentum \emph{wave action} can be derived \citep{1968RSPSA.302..529B, 1978JFM....89..647A}, given that the mean flow can be meaningfully defined. A similar application regarding spiral density waves in galactic disk was discussed in \citet{1969ApJ...158..899T} and \citet{1970ApJ...160...99S}. In the absence of self-gravity, we only need to consider the flux of wave action due to advective transport. In any case, as the radial flux of angular momentum wave action is conserved in the absence of dissipation, it can be used to relate the amplitude of the waves at different radii in the linear theory. For single-mode WKB (radial) waves (e.g., $P_1 \propto \exp{i\int k(r) dr}$), the time-averaged, vertically integrated flux of angular momentum wave action can be derived from Equations \eqref{eq:full3_2} and \eqref{eq:full3_3} and is given by \citep[e.g.,][]{1998ApJ...504..983L}
\begin{align}
F^{(a)} = \frac{\pi r m}{k} \left(\frac{\hat{\omega}^2 - \kappa^2}{\hat{\omega}^2}\right) \int \rho |u|^2 dz,
\end{align}
where $k$ is the radial wavenumber. Note that this equation is general for WKB theories with $k^2 \gg m^2/r^2$ (i.e., valid away from LRs) and is derived without our previous assumption in Section \ref{sec:linpert} to neglect azimuthal pressure gradient when computing the eigenfunctions. By expressing the above formula in terms of dimensionless variables, we have
\begin{align}
\label{eq:waveaction}
F^{(a)} &= \pi m \left[ r \frac{K \Omega^2 H^4 \rho_{\rm mid}(r)}{F^2(F^2-\kappa^2)} \right]|a|^2 \int^{Z_{\rm b}}_0 \rho(Z) X^2(Z) dZ,
\end{align}
where $\rho_{\rm mid}(r)$ is the midplane gas density and $\kappa=1$ is the normalized epicyclic frequency for a Keplerian disk. Here we denote $a=a(r)$ as the slowly-varying amplitude of the waves (i.e., $u(r,z) = a \tilde{u}(z)e^{ikr}$ at non-LR locations). On the left-hand-side of Equation \eqref{eq:waveaction}, the value of $F^{(a)}$ for each mode is determined by its corresponding value at the LRs \citep{1998ApJ...504..983L}, i.e., $F^{(a)}_j = f_j \mathscr{T}$ where $\mathscr{T}$ and $f_j$ are the fraction and total value of the torque in Equations \eqref{eq:r1} and \eqref{eq:r3}, respectively. 

The confinement and wave-channeling of the waves are examined by considering the density of angular momentum wave action, which is given by
\begin{align}
\mathscr{A}^{(a)} = \frac{m}{2\hat{\omega}}\rho(|u|^2 + |{\rm w}|^2),
\end{align}
where averaging over one wave period is assumed. Consequently, the vertically-integrated flux $A^{(a)} = \int^\infty_0 \mathscr{A}^{(a)} dz$, is related to the radial flux by $F^{(a)} = A^{(a)} {\rm v}_g$, where ${\rm v}_g = \partial \hat{\omega} / \partial k$ is the radial group velocity. For a monochromatic wave ($\omega/m=\Omega_{\rm p}$ being a constant), the relevant conservation law for the wave action reads
\begin{align}
\label{eq:conserve_waveaction}
\frac{dF^{(a)}}{dr}= 0,
\end{align}
which implies $F^{(a)}$ being a constant except at the LRs.

\begin{figure*}[!htb]
\centering
\includegraphics[width=0.33\textwidth]{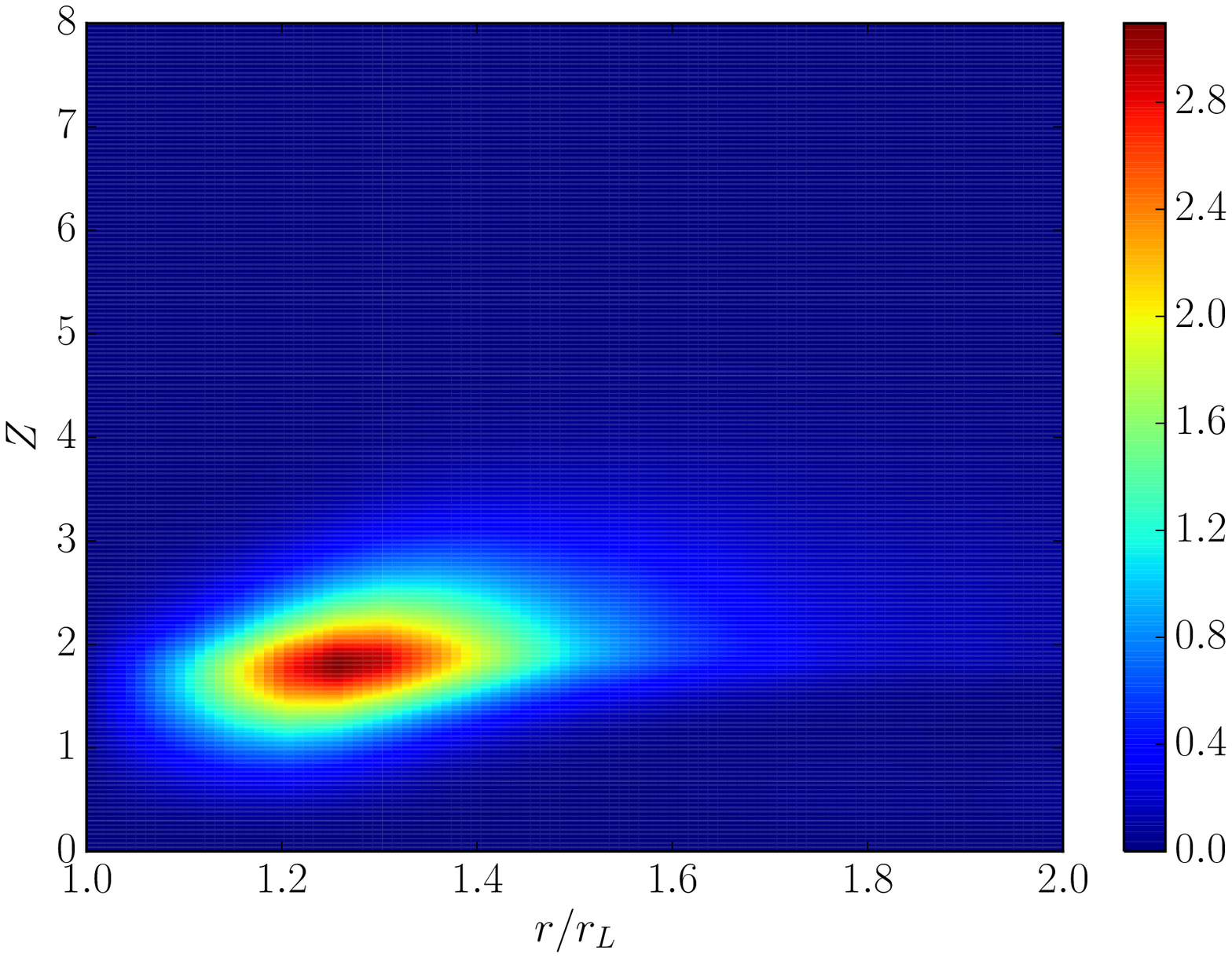}
\includegraphics[width=0.33\textwidth]{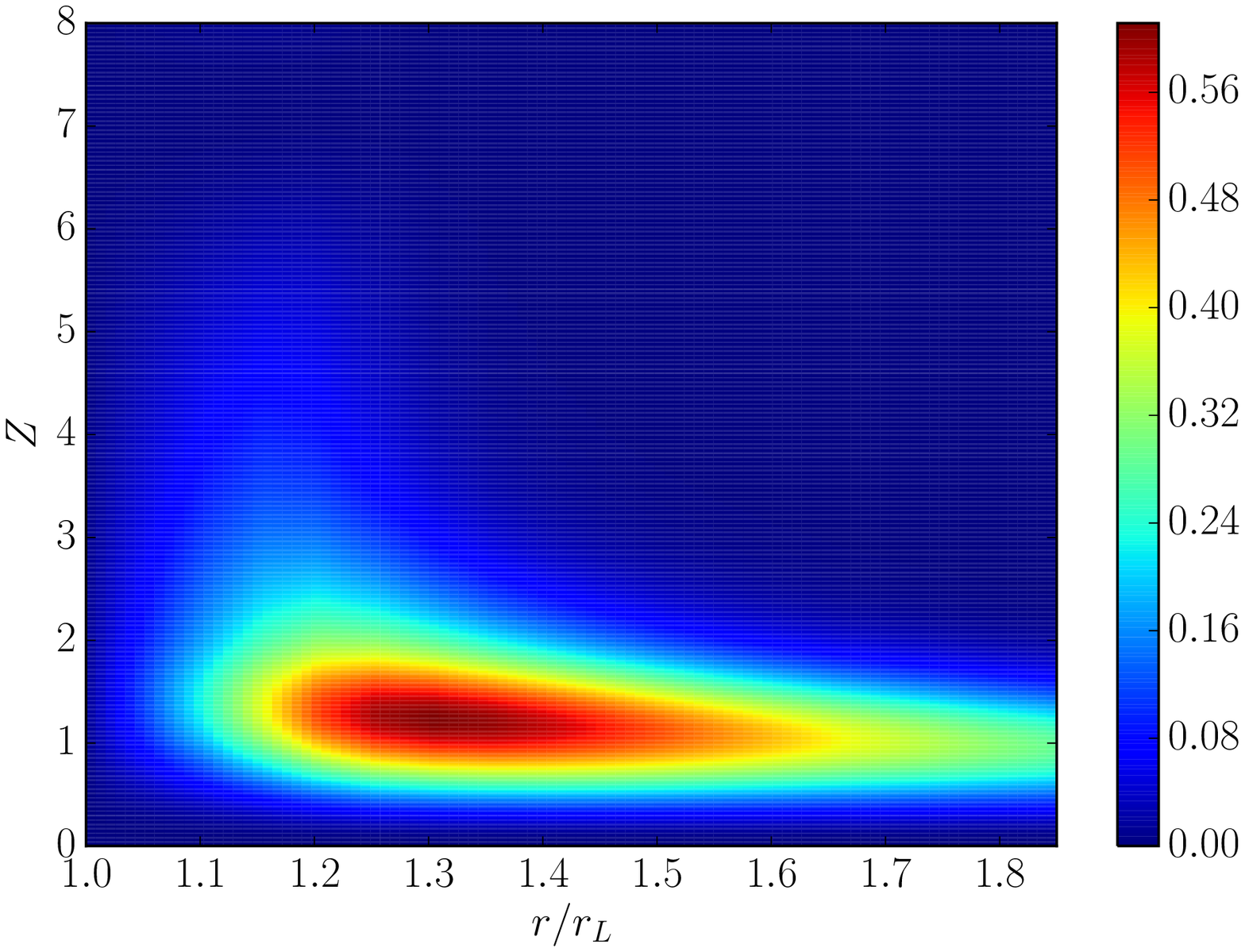}
\includegraphics[width=0.33\textwidth]{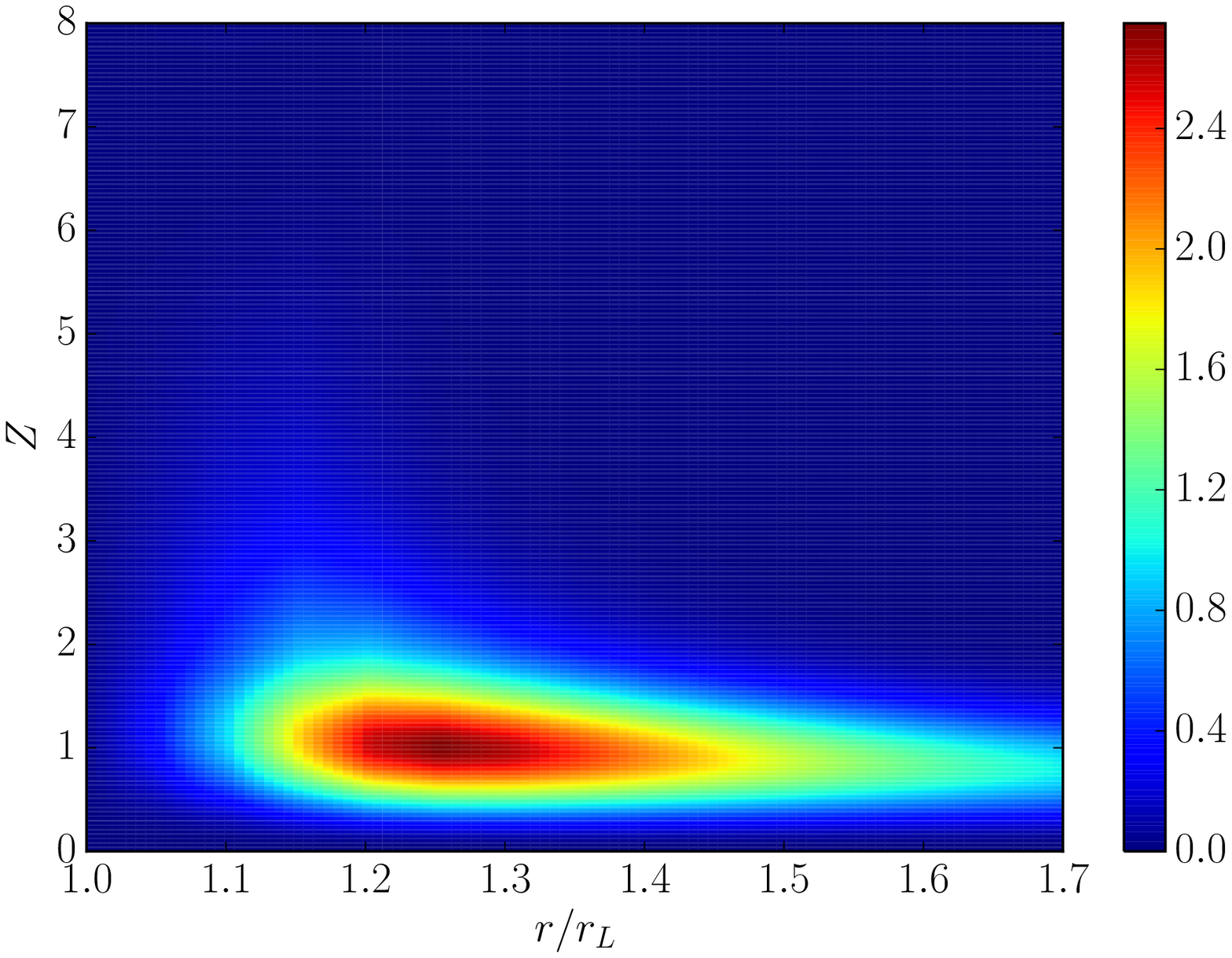}
\caption{Variations of density of angular momentum wave action for $f$-mode. The linear color scale in the arbitrary on each plot. The values of $A$ are $-0.5$, 0.5, and 2.0, for the left, middle, and right figures, respectively. The horizontal axis is between $r_L$ and $2r_L$.}
\label{fig:E-f-con}
\end{figure*}

\begin{figure*}[!htb]
\centering
\includegraphics[width=0.33\textwidth]{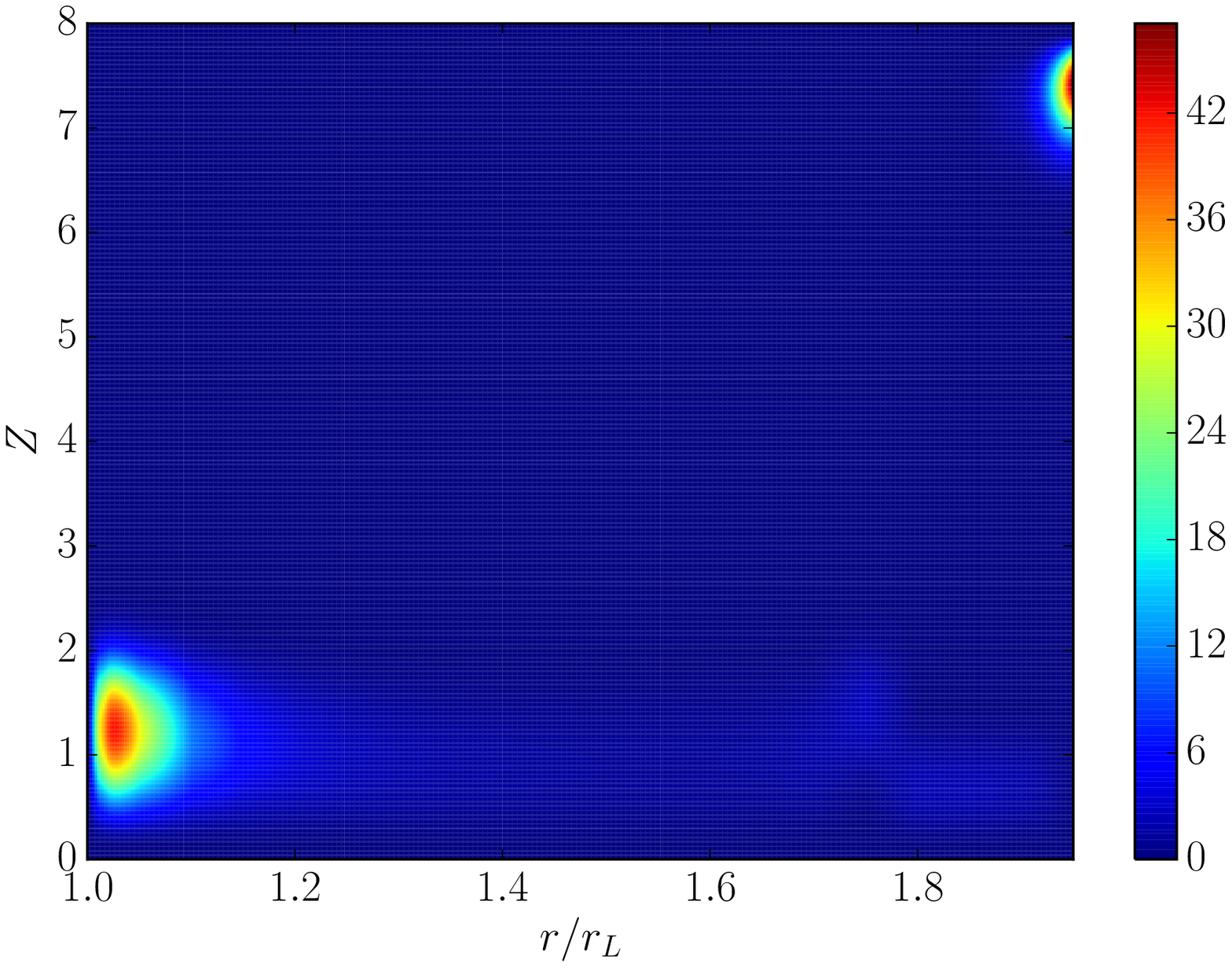}
\includegraphics[width=0.33\textwidth]{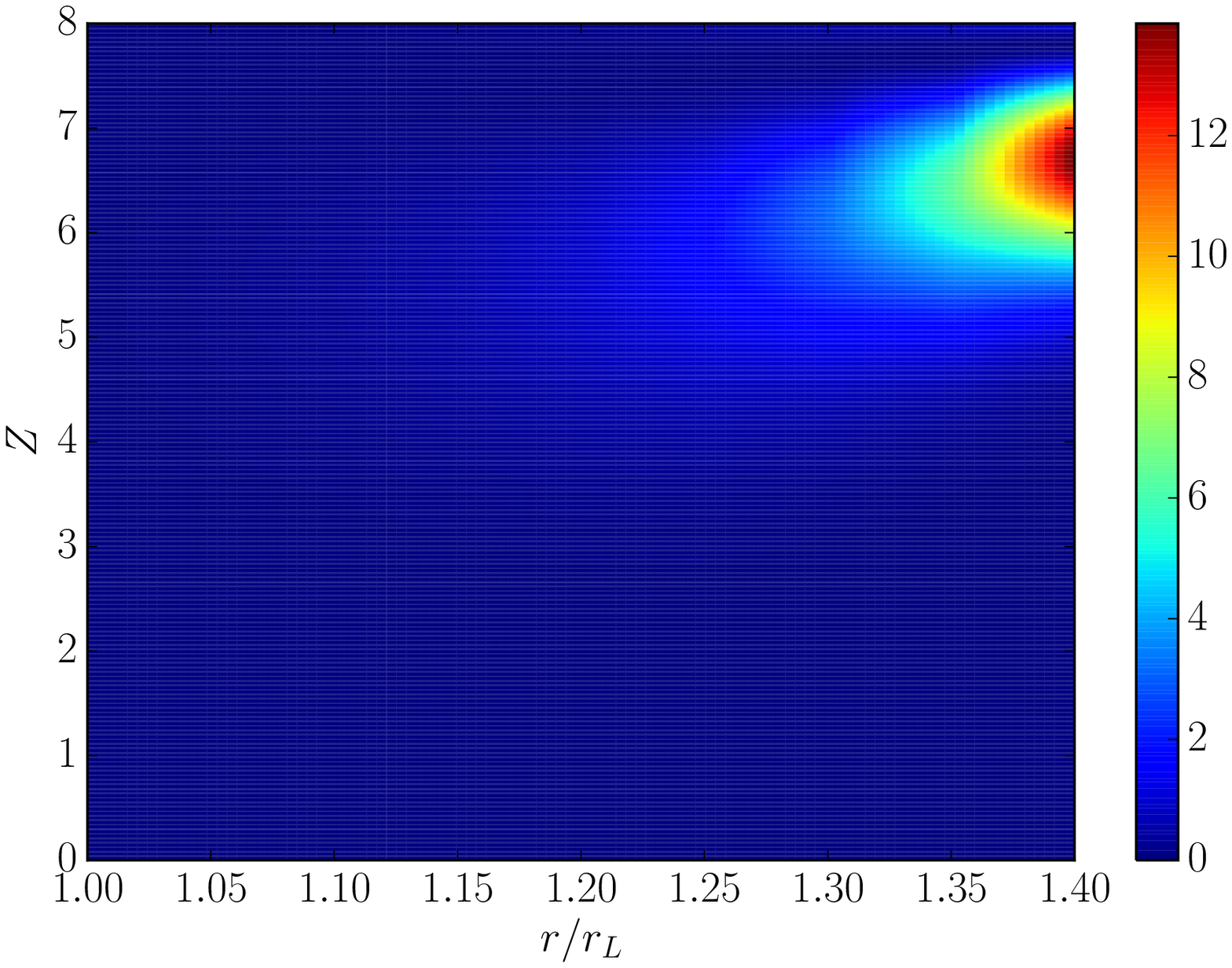}
\includegraphics[width=0.33\textwidth]{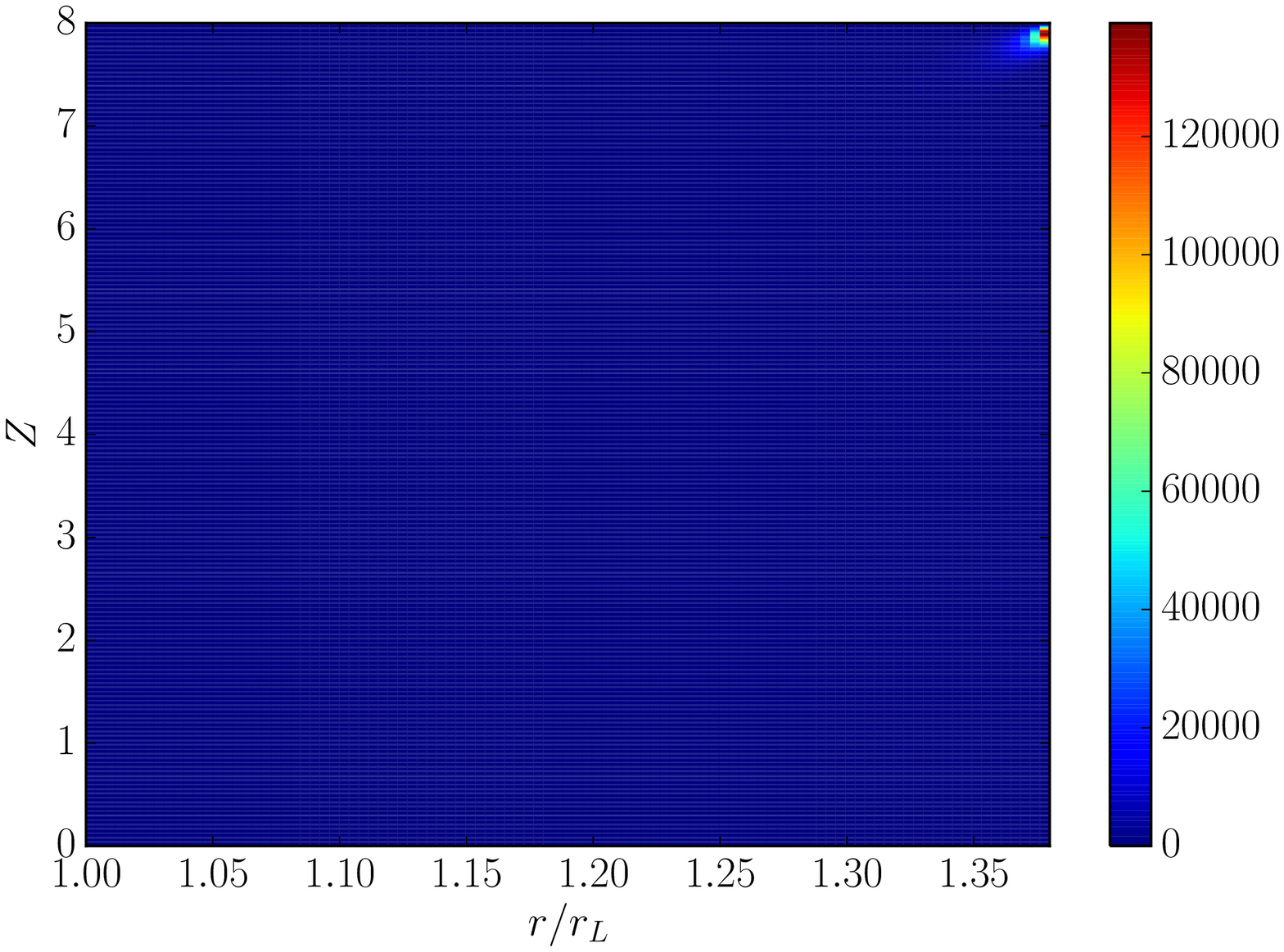}
\caption{Same as Figure \ref{fig:E-f-con}, but for $g_1$-mode.}
\label{fig:E-g1-con}
\end{figure*}

For a given disk model and a dispersion relation of a particular mode, the wave amplitude can be expressed as a function of $r$, which is obtained by keeping $F^{(a)}$ a constant and using normalization such as $\int\rho_0 X^2 dZ=1$ at \textit{each radius}. Here, we use simple power laws for midplane density ($\rho_{\rm mid}(r) \propto r^{-p}$) and temperature ($T_{\rm mid}(r) \propto r^{-q}$). In the following, $p=2.25$ and $q=1/2$ such that the surface density goes as $\Sigma_0 \propto \rho_0 H \propto r^{-1}$. In Figure \ref{fig:E-f-con}, the density of wave action $\mathscr{A}^{(a)}$ of $f$-mode are shown on the $r$-$z$ plane for $A=-0.5$, 0.5, and 2.0 and in a radial range of $r_L < r < 2r_L$. When the atmosphere is colder than the midplane ($A<0$), the density of wave action peaks at the midplane and gradually channels near the transition where $Z_{\rm a}=2.0$. This wave-channeling effect was also reported by \citet{1999ApJ...515..767O} where an isothermal atmosphere is attached to a polytropic disk near the midplane. Since the sound speed is continuous in our model across $Z_{\rm a}$, the density of wave action do not stay at a particular height as in their work. For the cases with a hotter atmosphere ($A>0$), the density of wave action peaks near (but not exactly at) the midplane. Instead of channeling to the base of hot atmosphere in the disk, the $f$-mode channels towards the midplane where the sound speed is lowest. This suggests that the direction of wave-channeling for $f$-mode depends on the temperature gradient, which is expected from its acoustic nature. For $g_1$-mode shown in Figure \ref{fig:E-g1-con}, $\mathscr{A}^{(a)}$  channels even higher as it propagates away from the LR and peaks close to the upper boundary.

While the vertical velocity perturbation $\tilde{w}(z)$ peaks at the upper boundary for $f$- and $g$-modes, the peak height of the density of angular momentum wave action differs. The concentration of wave energy locates at different heights due to the wave-channeling. Although the nonlinear effect is not included in the current work, the wave action density shown in Figures \ref{fig:E-f-con} and \ref{fig:E-g1-con} indicates the differences among different modes.

The upper boundary serves as a lid which allows surface modes to exist. This is only an idealization such that the waveform does not extend to infinity. This is expected that, far above the midplane, the thin-disk approximation breaks down and the thermal property of the disk changes.  At a greater height, where the heat from stellar irradiation can be easily radiated away, the thermal time is expected to be so short that the gas response becomes isothermal. This case was previously studied by \citet{1999ApJ...515..767O} where the isothermal gas response ($\gamma=1$) in the atmosphere is considered and a matching condition can be derived at the base of this layer. When such boundary condition is adopted (with a corresponding matching condition), no $g$-mode is found using our method. This is consistent with the speculation that $g$-modes are in the continuous spectrum.

Among the lower-order modes considered in this work, the eigenfunctions are mostly evanescent in the vertical direction. The vertical wavenumber is thus not well-defined. The vertical group velocity is essentially zero even though the number of nodes may change slightly (e.g., from zero to one) during propagation. The presence of wave-channeling instead of wave propagation in the vertical direction (i.e., refraction) was also confirmed in the asymmetrical hydrodynamic simulations in \citet{2002MNRAS.332..575B} for a polytropic disk.

\section{Summary and Conclusion}
\label{sec:conclusion}

In this work we explore the properties of the waves that are excited at the LRs in a planet-disk system. In particular, the disk models with a hotter atmosphere is studied. The linear theory under the WKB approximation is developed based on the earlier work by Lubow and Pringle and the separation of variables in terms of radial and vertical coordinates is applied. This model is essentially an 1+1D calculation, in which the eigenfunctions are calculated at each radius $r$ (or equivalently at each $\hat{\omega}$) and are connected depending on the radial structure of the disk by means of the conservation of wave action (see, Equation \ref{eq:conserve_waveaction}).

At each radius (including LRs), the local boundary value problem is solved for the eigenvalues and eigenfunctions. Each eigenvalue corresponds to different mode, which can be categorized based on their behavior for large radial wavenumber ($k$) \citep{1998MNRAS.297..291O}. For example, the linear dispersion relation of $f$-mode for large $k$ reflects its acoustic nature. In Section \ref{sec:torquefraction}, the eigenfunctions are used to compute the torque fraction of each mode. Unless $A$ is very large, the $f$-mode carries most of the angular momentum flux away. The temperature of the disk atmosphere and its gradient affect the background density and sound speed, which in turns change the contribution of different waves excited at LRs. As the wave propagates away from the LRs, the density of the wave action concentrates at different height. This wave-channeling phenomenon in the vertical direction differs from the usual wave propagation as the group velocity is only tangential to the disk plane (i.e., $\partial \hat{\omega}/\partial k_z = 0$).

For a disk atmosphere hotter than the midplane ($A>0$), we find that the $f$-mode remains confined within the layer closer to the midplane (Figure \ref{fig:E-f-con}), whereas the $g$-modes retain its surface-wave character and continue to channel upward (Figure \ref{fig:E-g1-con}). Since the group velocity of $f$-mode is almost constant when it propagates far enough from the LR, the concentration of $\mathscr{A}^{(a)}$ is mainly due to the change in eigenfunctions. On the other hand, as a result of the angular momentum conservation and the decrease in group velocity (Figure \ref{fig:KF}), the amplitudes of the $g$- and $r$-modes increase as they propagate. Depending on the strength of the tidal forcing (i.e., mass of the planet), the increasing amplitude may lead to nonlinear dissipation. Combining the effect of wave-channeling, our understanding on the formation and structure of the spiral shock may change for a three-dimensional disk.

We parameterize the vertical temperature profile $T(Z)$ for an equilibrium disk using Equation \eqref{eq:temperature1}, in order to mimic the results from previous radiative transfer calculations \citep[e.g.,][]{2002A&A...389..464D}. By expressing the governing equations in terms of the arbitrary temperature profile $T(Z)$, one can easily apply our formulation to study different models used in observations and simulations.
However, identification of different modes for each $m$ in a realistic planet-disk system remains a challenge and will be left for future investigation. In conclusion, we study the behaviors of different waves that are excited at LRs. The properties of these linear waves will help us better understand the tidal interactions between planet and disk.

\subsection{Application to Nonlinear Effects}

\begin{figure*}[!ht]
\centering
\includegraphics[width=0.33\textwidth]{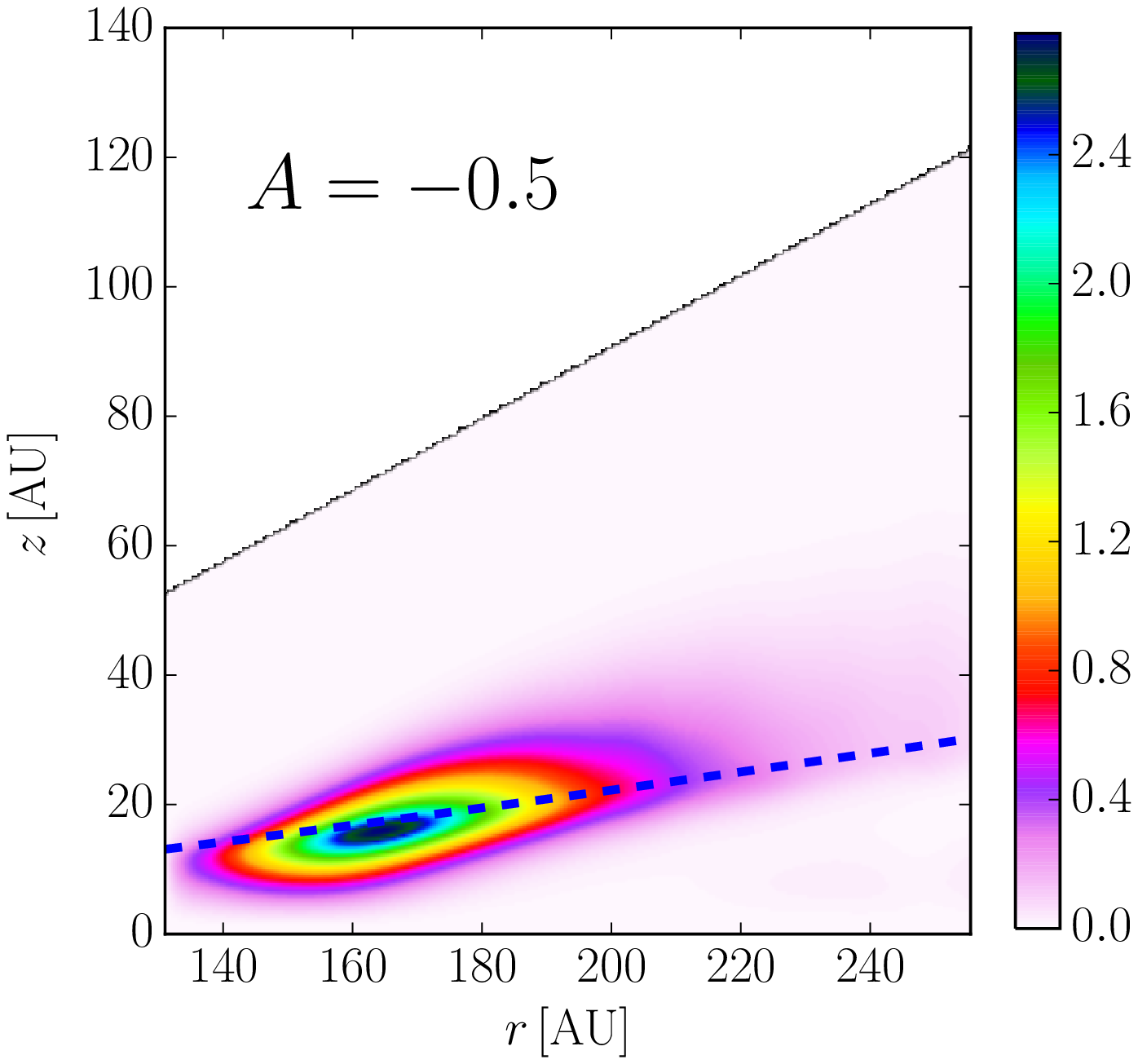}
\includegraphics[width=0.33\textwidth]{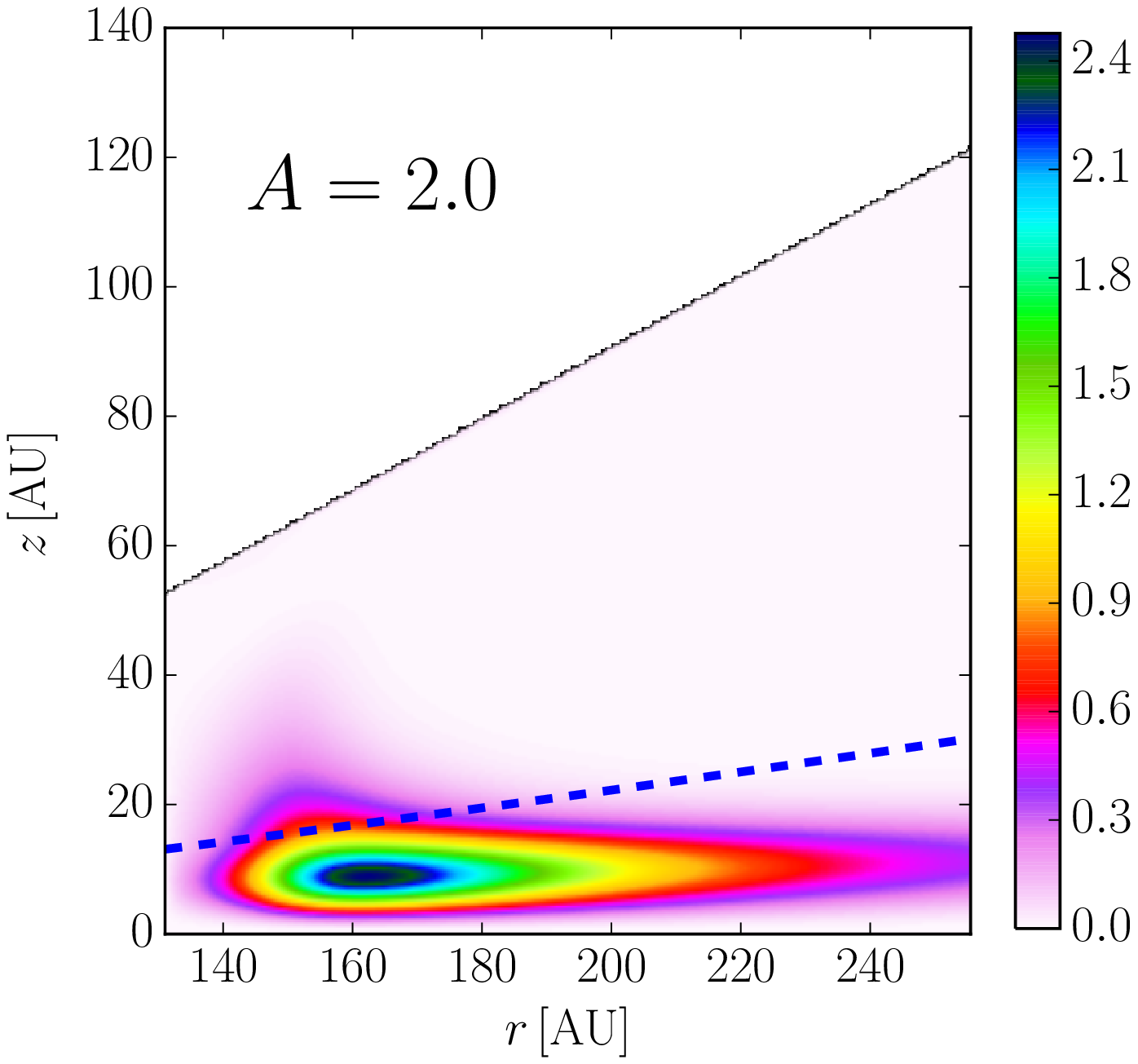}
\caption{Variations of density of angular momentum wave action $\mathscr{A}^{(a)}$ for $f$-mode, in physical space. The linear color scale in the arbitrary on each plot. The values of $A$ are $-0.5$ and 2.0, for the left and right panels, respectively. The blue dotted line indicates the transition height at $z=2H(r)$. The horizontal axis is between $r_L$ and $2r_L$ for $m=2$. The corotation radius is at 100$\,{\rm AU}$. The power-law index for density and temperature are $p=2.25$ and $q=0.5$, respectively.}
\label{fig:E-f-con-wedge}
\end{figure*}

Lastly, the linear study of waves may still give us some hints about the nonlinear processes. There are a few factors determining whether the waves will eventually break, for example, the strength of the forcing, local sound speed, and the character of the mode. For $f$-mode, the wave energy are confined within the central colder layer, but the amplifications are only moderate and not very sensitive to the choice of $A$ (see Figure \ref{fig:E-f-con-wedge}). For $g$-modes, which carry more angular momentum in the case of hotter disk atmosphere, they amplifies significantly and channel into the top of disk atmosphere. It will be interesting to investigate the nonlinear behavior of $g$-modes in the atmosphere. Other dissipation processes such as viscous damping \citep{1996ApJ...460..832T} and radiative damping \citep{1996ApJ...472..789C} may be relevant to the linear waves in this work as well.

Together with the finding that the torque fraction carried by each mode depends sensitively on the temperature gradient, we expect that the shock structure (e.g., shock location) in such a three-dimensional disk may be very different from the two-dimensional disk. Consequently, the change in the location of the deposit of angular momentum may lead to differences in terms of planet migration as well. In particular, the excitation of $g$-modes in a thermal-stratified disk may modify what we expect from an isothermal disk in the literature \citep[e.g.,][]{2002ApJ...565.1257T}.

Nevertheless, recent three-dimensional numerical simulations of planet-disk interaction by \citet{2015arXiv150703599Z} show that the inner spiral shock (within corotation radius) are formed at a few scale height from the midplane (but not the computational boundary). While the wave-channeling effect in an initially-isothermal disk adopted in their study is expected to be insignificant, the nonlinear effects from the shock-heating at the surface is unknown. Therefore, our study on the wave channeling may provide a basis for future investigations. 

The work is supported by a MOST grant in Taiwan through MOST 103-2112-M-001-027.

\appendix
\section{A. Away from the Lindblad Resonances}
\label{sec:appendixA}
Away from the (horizontal) resonance locations where the waves are excited, the radial wavelength of the perturbation is small compared to the variation length scale of the background equilibrium. The perturbational equations are very similar to the case near LRs, except there is no explicit $x$ dependence in Equation \eqref{eq:full3_6}. \citet{1993ApJ...409..360L} and \citet{1995MNRAS.272..618K} performed the local calculations of the axisymmetric $(m=0)$ free waves in the vertically isothermal and polytropic disks, respectively. In order to perform the normal mode analysis for non-axisymmetric waves, we further neglect the advection terms due to the background shear as done in \citet{1998ApJ...504..983L}. One reason is that the mixture of time ($t$) and space ($r$) does not allow the normal modes to exist in the form of shearing waves \citep{1992ApJ...388..438R}. Since we are more interested in the vertical structure of waves at different radii, the information on the radial front is sacrificed. In this limit of WKB approximation in the radial direction, we apply the Fourier transformation in $r$ for the perturbational variables (i.e., $\exp{ikr}$). By keeping $\hat{\omega}(r)$ as a local constant, we effectively dropped the differential shear in the disk \citep{1989ApJ...347..435V}. Therefore, using the dimensionless variables, the differential equations become
\begin{align}
\label{eq:dxdzNLR}
\frac{dX}{dZ} &= \frac{N^2}{Z} X + (F^2 - N^2) W, \\
\label{eq:dwdzNLR}
\frac{dW}{dZ} &= \left[\frac{K^2}{F^2-\kappa^2} - \frac{\rho(Z)}{\gamma P(Z)}\right] X - \frac{P'(Z)}{\gamma P(Z)} W,
\end{align}
where we define
\begin{align}
X(Z) &= -\frac{i\hat{\omega}\tilde{P}_1(z)}{\Omega^3 H^2 \rho_0}, \quad W(Z) = \frac{\tilde{w}(z)}{\Omega H}, \quad K = kH, \\
\end{align}
and $H$ is the scale height at the midplane (Equation \eqref{eq:midplanescaleheight}). In terms of $T(Z)$, we have
\begin{align}
\label{eq:dxdzNLR_T}
\frac{dX}{dZ} &= \left[\left(\frac{\gamma-1}{\gamma}\right)\frac{Z}{T(Z)}+\frac{T'(Z)}{T(Z)}\right]X + \left[F^2 - \left(\frac{\gamma-1}{\gamma}\right)\frac{Z^2}{T(Z)}-\frac{ZT'(Z)}{T(Z)}\right] W, \\
\label{eq:dwdzNLR_T}
\frac{dW}{dZ} &= \left[\frac{K^2}{F^2-\kappa^2} - \frac{1}{\gamma T(Z)}\right] X + \frac{Z}{\gamma T(Z)} W.
\end{align}

\section{B. Vertical Hydrostatic Equilibrium}
\label{sec:appendixB}

For $n=2$ in Equation \eqref{eq:temperature1}, the vertical hydrostatic equilibrium can be solved analytically. The dimensionless equilibrium pressure is given by
\begin{align}
P(Z) = \begin{cases}
\exp\left\lbrace-\frac{Z_{\rm a}^2}{2\sqrt{1+A}}\arctan\left[\sqrt{1+A}\tan \pi \left(\frac{Z}{Z_{\rm a}}\right)^2 \right]\right\rbrace & \mbox{if } Z < Z_{\rm a}\\
\exp\left[-\frac{Z_{\rm a}^2}{2\sqrt{1+A}}-\frac{Z^2-Z_{\rm a}^2}{2(1+A)}\right] &\mbox{if } Z \geq Z_{\rm a}, \\ 
\end{cases}
\end{align}
where $Z_{\rm a}$ is the transition height. The equilibrium density is given by $\rho(Z) = P(Z)/T(Z)$.

\bibliography{diskwaves}
\end{document}